\documentclass[prd,showpacs,preprintnumbers,amsmath,amssymb]{revtex4}

\usepackage{graphicx}
\usepackage{dcolumn}
\usepackage{bm}

\begin{document}

\title{Exact Cosmological Solutions of 4D String Gravity in the String Frame}

\author{Yongsung Yoon}\email{cem@hanyang.ac.kr}
\affiliation{Physics Department, Research Institute for Natural Sciences, Hanyang University, Seoul 133-791, Korea}

\begin{abstract}
We have discussed a particular class of exact cosmological solutions of the 4-dimensional low energy string gravity in
the string frame. In the vacuum without matter and the 2-form fields, the exact cosmological solutions always give
monotonically shrinking universes if the dilaton field is not a constant. However, in the presence of the 2-form fields
and/or the radiation-like fluid in the string frame, the exact cosmological solutions show a minimum size of the
universe in the evolution, but with an initial cosmological curvature singularity in the string frame.
\end{abstract}
 \maketitle

\section{Introduction}

The low energy limit of string theory provides a definite form of dilaton gravity action \cite{IGO1,IGO2,IGO3,IGO4}
with a specific dilaton coupling and additional 2-form fields \cite{GSW}. Thus, it may be worth examining the exact
cosmological solutions of the 4-dimensional low energy string gravity in the framework of dilaton gravity, instead of
General Relativity.

There have been many studies on the cosmological evolution of the low energy string gravity with 2-form fields
\cite{intro,intro2,scos1,scos2,scos3}. In the early cosmological evolution, it is necessary to consider the effect of
matter as well as the 2-form fields in the low energy string gravity. Without matter, it would be convenient to choose
the Einstein frame redefining the metric as usual. However, in the presence of matter, the frame changes of metric make
the dilaton field couple with matter fields differently depending on their spins. Thus, the matter energy-momentum
tensors would be non-trivial functions of the dilaton field in a different frame.

Therefore, we have considered a fluid-like matter distribution in the fundamental string frame \cite{frame}, in which
the gravitational constant determined by the dilaton field evolves in time. Because the relativistic matter should be
dominant over the non-relativistic matter in the early evolution of the universe, we have examined the exact
cosmological solutions with the 2-form fields and/or the radiation-like fluid in the string frame.

\section{Low Energy String Gravity in the String Frame}

The vanishing beta functions originated from the world sheet conformal symmetry in the closed string sector of string
theory give the well-known low energy string gravity action in the critical $D$-dimension \cite{action}. Assuming a
Ricci-flat compactification of the internal $(D-4)$-dimensional space decoupled from our $4$-dimensional spacetime
\cite{CY,witten}, and adding the matter Lagrangian ${\cal L}_{m}$ which is decoupled from the dilaton field in the
string frame, we have the following form of action in 4-dimensional spacetime from the low energy limit of string
theory \cite{GSW,action,string,D-brane,gasperini},
\begin{eqnarray}
S= -\frac{1}{2}\int d^{4}x\sqrt{-g}\left[\Phi^{2} R +4(\partial \Phi)^{2}
-\frac{1}{12}\Phi^{2}H_{\alpha\mu\nu}H^{\alpha\mu\nu} \right] + \int d^{4}x\sqrt{-g} {\cal L}_{m} , \label{action}
\end{eqnarray}
where we have redefined the dilaton field $\phi$ with $\Phi$ as
\begin{eqnarray}
\Phi \equiv \frac{1}{\kappa_{0}} e^{-\phi} \label{redef-eq}
\end{eqnarray}
in the string frame, and $H_{\alpha\mu\nu}$ are the field strengths of the 2-form antisymmetric fields $B_{\mu\nu}$ in
NS-NS sector as ${\bf H}=d{\bf B}$.

The equations of motion for the 2-form fields $B_{\mu\nu}$, the dilaton field $\Phi$, and the metric $g_{\mu\nu}$ are
obtained as
\begin{eqnarray}
\nabla_{\alpha}(\Phi^{2}H^{\alpha\mu\nu}) = \frac{1}{\sqrt{-g}}\partial_{\alpha}(\sqrt{-g}\Phi^{2}H^{\alpha\mu\nu}) =0
, \label{H-eq}
\end{eqnarray}
\begin{eqnarray}
4\nabla^{2}\Phi = \Phi R -\frac{1}{12}\Phi H_{\alpha\mu\nu}H^{\alpha\mu\nu} , \label{D-eq}
\end{eqnarray}
\begin{eqnarray}
\Phi^{2} G_{\mu\nu} - 2\Phi\nabla_{\mu}\nabla_{\nu}\Phi +2g_{\mu\nu}\Phi\nabla^{2}\Phi +2\nabla_{\mu}\Phi
\nabla_{\nu}\Phi -\frac{1}{4}\Phi^{2}H_{\mu\alpha\beta}H_{\nu}^{\alpha\beta}
+\frac{1}{24}g_{\mu\nu}\Phi^{2}H_{\alpha\beta\lambda}H^{\alpha\beta\lambda} = T^{(m)}_{\mu\nu} , \label{G-eq}
\end{eqnarray}
where $T^{(m)}_{\mu\nu}$ is the matter energy-momentum tensor. Taking the trace part of equation (\ref{G-eq}), we have
an expression for $\Phi^{2}R$. Plugging it into the dilaton equation (\ref{D-eq}), we can rewrite the dilaton equation
of motion equivalently as
\begin{eqnarray}
2\left(\Phi \nabla^{2}\Phi + (\nabla\Phi)^{2}\right) = \frac{1}{6}\Phi^{2}H_{\alpha\mu\nu}H^{\alpha\mu\nu} + T^{(m)}
\label{S-eq},
\end{eqnarray}
where $T^{(m)}$ is the trace of the matter energy-momentum tensor.

As we can see from Eq.(\ref{S-eq}), the driving force of the dilaton field is given by the antisymmetric 2-form field
strengths and the trace of matter energy momentum tensor.

$H_{\alpha\mu\nu}$ solution of Eq.(\ref{H-eq}) is found as
\begin{eqnarray}
H^{\alpha\mu\nu}=\Phi^{-2}\epsilon^{\alpha\mu\nu\lambda}\partial_{\lambda}\psi , \label{H-sol}
\end{eqnarray}
where $\psi$ is a new smooth scalar field, $\epsilon^{\alpha\mu\nu\lambda}$ is
the covariant constant antisymmetric tensor such that
$\epsilon^{\alpha\mu\nu\lambda}=\varepsilon^{\alpha\mu\nu\lambda}/\sqrt{-g}$,
with $\varepsilon^{0123}= +1$. The Bianchi identity $d{\bf H}=0$ for the
antisymmetric field strengths $H_{\alpha\mu\nu}$ gives the equation of motion
for the scalar field $\psi$ as \cite{gasperini,known1,known2,scos1,scos2,scos3}
\begin{eqnarray}
\nabla^{\alpha}(\Phi^{-2}\partial_{\alpha}\psi)=0 . \label{psi-eq}
\end{eqnarray}

Defining the dual strength vector $h_{\alpha}$ as $H^{\mu\nu\beta} \equiv \epsilon^{\mu\nu\beta\alpha}h_{\alpha}$, we
find that
\begin{eqnarray}
\frac{1}{12}H_{\alpha\mu\nu}H^{\alpha\mu\nu} =-\frac{1}{2}\Phi^{-4}(\nabla\psi)^{2}=-\frac{1}{2}h^{\alpha}h_{\alpha} .
\label{H2-sol}
\end{eqnarray}

We examine the cosmological evolution of the Eqs.(\ref{H-eq},\ref{D-eq},\ref{G-eq}) with the spatially flat
Robertson-Walker metric
\begin{eqnarray}
ds^{2}=dt^{2}-S^{2}(t)d\vec{\bf r}^{2}, \label{RW}
\end{eqnarray}
and the cosmological Ansatz: $\Phi=\Phi(t)$, $\psi=\psi(t)$. Then, Eq.(\ref{psi-eq}) gives
\begin{eqnarray}
\frac{d\psi}{dt}=\Phi^{2}\frac{A}{S^{3}} , \quad {\rm i.e.} ~H^{ijk}=\epsilon^{ijk0}\frac{A}{S^{3}}, \quad {\rm and}
~H^{0ij}=0 , \label {h-sol2}
\end{eqnarray}
where $A$ is a constant which measures the strength of the 2-form fields. The solution Eq.(\ref{h-sol2}) gives a
uniform non-vanishing dual strength vector such as
\begin{eqnarray}
h_{i} = 0, \quad h_{0}=-\frac{A}{S^{3}} . \label{B-sol}
\end{eqnarray}

Using the metric (\ref{RW}) and the explicit solution of $H_{\mu\nu\lambda}$
from Eqs.(\ref{H2-sol},\ref{B-sol}), Eqs.(\ref{G-eq},\ref{S-eq}) are reduced to
a set of cosmological equations. Denoting the time derivative as an over-dot,
the cosmological equations with uniform matters in the string frame are given
as
\begin{eqnarray}
3\Phi^{2}{\cal H}^{2} +6{\cal H}\Phi\dot{\Phi} +2\dot{\Phi}^{2} = \rho_{m} +
\frac{1}{4}\Phi^{2}\frac{A^{2}}{S^{6}} , \label{CosG00-eq}
\end{eqnarray}
\begin{eqnarray}
2 {\bf a} \Phi^{2} +{\cal H}^{2}\Phi^{2} +4{\cal H}\Phi\dot{\Phi}
+2\Phi\ddot{\Phi} = -p_{m} - \frac{1}{4}\Phi^{2}\frac{A^{2}}{S^{6}} ,
\label{CosG11-eq}
\end{eqnarray}
\begin{eqnarray}
2(\Phi\ddot{\Phi} +3{\cal H}\Phi\dot{\Phi} +\dot{\Phi}^{2}) = \rho_{m} -3p_{m}
- \Phi^{2}\frac{A^{2}}{S^{6}} , \label{CosD-eq}
\end{eqnarray}
where the Hubble parameter is ${\cal H} \equiv \dot{S}/S$, and the cosmic
acceleration parameter is ${\bf a} \equiv \ddot{S}/S$. With the uniform matter
density $\rho_{m}$ and the uniform matter pressure $p_{m}$, the energy momentum
tensors are $~T^{(m)0}_{0}=\rho_{m}$, and
$~T^{(m)1}_{1}=T^{(m)2}_{2}=T^{(m)3}_{3}=-p_{m}$.

In Eqs.(\ref{CosG00-eq},\ref{CosG11-eq}), it seems that the 2-form fields play
a role of matter with the energy density $\Phi^{2}\frac{A^{2}}{4S^{6}}$ and the
pressure $\Phi^{2}\frac{A^{2}}{4S^{6}}$ having the equation of state $w=1$.
However, the $\Phi^{2}H_{\alpha\mu\nu}H^{\alpha\mu\nu}$ term in the action
(\ref{action}) gives an additional contribution, the conformal density
$-\Phi^{2}\frac{A^{2}}{2S^{6}}$ defined as $\Phi\partial{\cal L}/\partial\Phi$,
to the right hand side of Eq.(\ref{CosD-eq}).

By taking a time derivative Eq.(\ref{CosG00-eq}) and using
Eqs.(\ref{CosG11-eq},\ref{CosD-eq}), it is found that the matter energy
conservation holds as
\begin{eqnarray}
\frac{d}{dt}\rho_{m} + 3H(\rho_{m}+p_{m}) = 0 , \label{matter-eq}
\end{eqnarray}
which is independent of the 2-from fields $B_{\mu\nu}$ and the dilaton field $\Phi$ in the string frame. With the
equation of state for matter $w$, Eq.(\ref{matter-eq}) gives the matter density evolving as
\begin{eqnarray}
\rho_{m}(t) = \frac{C}{S^{3(w+1)}} , \label{matter-sol}
\end{eqnarray}
where $C$ is a positive constant which measures the strength of the matter density.

The 2-form fields having $\frac{A^{2}}{S^{6}}$ behavior in
Eqs.(\ref{CosG00-eq},\ref{CosG11-eq},\ref{CosD-eq}) dominate the universe with
a small scale factor $S(t)$. As the universe expands, the rapidly decreasing
2-form fields lay down their role in the late time evolution of the universe,
and matter with the density (\ref{matter-sol}) or a cosmological constant will
take over the dominance.

\section{The Exact Vacuum Solutions}

First, let us begin with the vacuum solution of Eqs.(\ref{CosG00-eq},\ref{CosG11-eq},\ref{CosD-eq}), without the 2-form
fields and matter. The equation (\ref{CosD-eq}) can be written as
\begin{eqnarray}
S^{3}\frac{d}{dt}\left(S^{3}\frac{d}{dt}\Phi^{2}\right) = 0 . \label{phi-vac-eq}
\end{eqnarray}
With a new time function $\tau(t)$ such that $d\tau \equiv dt/S^{3}(t)$, this equation becomes a simple equation having
a general linear solution in $\tau$,
\begin{eqnarray}
\frac{d^{2}}{d\tau^{2}}\Phi^{2} = 0, \quad \Phi^{2}=\alpha\tau+\beta \label{phi-vac-sol} ,
\end{eqnarray}
where $\alpha$ and $\beta$ are two positive integration constants which should
be determined by an initial condition. If $\alpha=0$, then we have the trivial
vacuum solution such that $\Phi^{2}=\beta$ is a constant, and the scale factor
$S$ is also a constant with ${\cal H}=0={\bf a}$. If $\alpha \neq 0$, then we
have
\begin{eqnarray}
\frac{\dot{\Phi}}{\Phi} = \frac{\alpha}{2S^{3}(\alpha\tau+\beta)} \label{phidot-vac-sol} .
\end{eqnarray}

The new time function $\tau$ is a monotonically increasing function of time $t$. As the time $t$ elapses, $\tau$
increases rapidly for a small scale factor, but very slowly for a large scale factor $S$. Thus, the gravitational
constant, which is the inverse of $\Phi^{2}$, decreases as $1/(\alpha\tau+\beta)$ in vacuum.

From Eq.(\ref{CosG00-eq}), we obtain $dS/d\tau$ and its integration as
\begin{eqnarray}
\frac{dS}{d\tau} = \eta\frac{\alpha S}{\alpha\tau+\beta} < 0, \quad
S=\sigma_{0}(\alpha\tau+\beta)^{\eta}, \quad \eta \equiv -\frac{\sqrt{3}\mp
1}{2\sqrt{3}} \label{scale-vac-eq} .
\end{eqnarray}
where $\sigma_{0}$ is a positive integration constant.

Thus, using $d\tau/dt=1/S^{3}(t)$, we obtain the cosmological time $t$ as a
function of $\tau$ as
\begin{eqnarray}
t - t_{0} = \frac{\sigma_{0}^{3}}{\alpha(3\eta+1)}(\alpha\tau+\beta)^{3\eta+1} \label{t-tau},
\end{eqnarray}
where $t_{0}$ is a constant time. Thus, we find the exact vacuum solutions without matter and the 2-form fields as
\begin{eqnarray}
S=\sigma_{0}\left(\frac{\alpha}{\sigma_{0}^{3}}(3\eta+1)(t-t_{0})\right)^{\eta/(3\eta+1)}, \quad
\Phi^{2}=\left(\frac{\alpha}{\sigma_{0}^{3}}(3\eta+1)(t-t_{0})\right)^{1/(3\eta+1)} \label{scale-vac-sol} ,
\end{eqnarray}
where $3\eta+1=-(1\mp\sqrt{3})/2$ and $\eta/(3\eta+1)=\mp 1/\sqrt{3}$.

Therefore, the both vacuum solutions with $\alpha \neq 0$ give an ever shrinking universe to $S=0$ in the future.
Especially, the solution with the lower signs meets a curvature singularity at a finite time $t=t_{0}$.

\section{The Exact Solutions with the 2-Form Fields only}

Let us consider the solution of Eqs.(\ref{CosG00-eq},\ref{CosG11-eq},\ref{CosD-eq}) with the 2-form fields only. The
equation (\ref{CosD-eq}) can be written as
\begin{eqnarray}
S^{3}\frac{d}{dt}\left(S^{3}\frac{d}{dt}\Phi^{2}\right) = -A^{2}\Phi^{2} \label{phi-eq}.
\end{eqnarray}
With a new time function $\tau(t)$ such that $d\tau \equiv dt/S^{3}(t)$, this equation becomes the simple harmonic
equation in $\tau$,
\begin{eqnarray}
\frac{d^{2}}{d\tau^{2}}\Phi^{2} = -A^{2}\Phi^{2}, \quad \Phi^{2}=\Phi^{2}_{0}\sin(A\tau), \quad \frac{\dot{\Phi}}{\Phi}
= \frac{A}{2S^{3}}\cot(A\tau) \label{AB-phi-sol} ,
\end{eqnarray}
where $\Phi_{0}$ is an integration constant, and $A \neq 0$. We find that $\tau$ should be in a compact domain, $0 \leq
\tau \leq \pi/A$.

From Eq.(\ref{CosG00-eq}), we have the solution of $H$ as follows,
\begin{eqnarray}
{\cal H}=-\frac{\dot{\Phi}}{\Phi} \pm
\frac{1}{\sqrt{3}}\sqrt{\left(\frac{\dot{\Phi}}{\Phi}\right)^{2}+\frac{A^{2}}{4S^{6}}} \label{Htau-sol} .
\end{eqnarray}
Using the solution (\ref{AB-phi-sol}) of $\Phi$, from Eq.(\ref{Htau-sol}), we find the velocity of the cosmic scale
factor as follows
\begin{eqnarray}
\frac{dS}{d\tau} = -\frac{AS}{2}\left(\cot(A\tau) \mp \frac{1}{\sqrt{3}}\csc(A\tau)\right) \label{AB-vel-eq} .
\end{eqnarray}
We can integrate Eq.(\ref{AB-vel-eq}) to have the scale factor $S$ as a function of $\tau$
\begin{eqnarray}
S_{\pm}(\tau)=S_{0}\frac{\tan^{\pm\frac{1}{2\sqrt{3}}}(A\tau/2)}{\sqrt{\sin(A\tau)}},
\end{eqnarray}
for the upper and the lower signs of Eq.(\ref{Htau-sol}) respectively.

The cosmological time $0 \leq t < \infty$ for each solution is given by the integration over $0 \leq \tau < \pi/A$,
\begin{eqnarray}
t(\tau)= \int_{0}^{\tau} S_{\pm}^{3}(\tau) d\tau \label{ctime} .
\end{eqnarray}
The infinite future of the cosmological time $t$ corresponds to $\tau=\pi/A$.

The cosmological time $t_{E}$ and the scale factor $S_{E}$ in the Einstein frame are obtained as
\begin{eqnarray}
t_{E}(\tau)= \int_{0}^{\tau} \Phi(\tau) S_{\pm}^{3}(\tau) d\tau, \quad S_{E}(\tau)= \Phi(\tau) S_{\pm}(\tau) =
|\Phi_{0}|S_{0}\tan^{\pm\frac{1}{2\sqrt{3}}}(A\tau/2) \label{Eframe} .
\end{eqnarray}
Because the time derivative of $S_{E}$ about $t_{E}$ is given as
\begin{eqnarray}
\frac{dS_{E}}{dt_{E}} = \pm \frac{A}{2\sqrt{3}} \frac{\csc(A\tau)}{S_{\pm}^{2}(\tau)} \label{Eframe2} ,
\end{eqnarray}
it is found that the upper/lower sign solution corresponds to an ever expanding/shrinking solution, respectively, in
the Einstein frame. The second derivative of $S_{E}$ about $t_{E}$ is the same for the both solutions, and always
negative as
\begin{eqnarray}
\frac{d^{2}S_{E}}{dt_{E}^{2}} = - \frac{A^{2}}{6|\Phi_{0}|}\frac{\csc^{5/2}(A\tau)}{S_{\pm}^{5}(\tau)} \label{Eframe2}
.
\end{eqnarray}
Thus, both solutions decelerate always in the Einstein frame.

The exact solutions in 4-spacetime dimension seem to be consistent with the phase-plane analysis \cite{scos2}, and with
the analytic Fourier mode solutions \cite{scos3}.

\section{The Exact Solutions with the Radiation-like Fluid Only in the String Frame}

Next, let us examine the cosmological solution of Eqs.(\ref{CosG00-eq},\ref{CosG11-eq},\ref{CosD-eq}) with the
radiation-like fluid in the string frame. Because the uniform relativistic matter has the equation of state $w=1/3$
with $p_{m}=w\rho_{m}$, the trace of the matter energy momentum tensors vanishes. Thus, without the 2-form fields
$A=0$, the dilaton equation (\ref{CosD-eq}) gives the same solution (\ref{phi-vac-sol}) as in the vacuum. If
$\alpha=0$, then we have the well-known decelerating expansion solution with a constant dilaton field $\Phi^{2}=\beta$
as in General Relativity.

Thus, let us consider the $\alpha \neq 0$ case, $\Phi^{2}=\alpha\tau+\beta$. Using a relativistic matter density
$\rho_{m}$ and a pressure density $p_{m}$ such that $\rho_{m}=3p_{m}=C/S^{4}$ from Eq.(\ref{matter-sol}), and with the
dilaton solution (\ref{phi-vac-sol},\ref{phidot-vac-sol}), we obtain the following equation for the scale factor $S$
from Eq.(\ref{CosG00-eq}),
\begin{eqnarray}
\frac{dS}{d\tau} = -\frac{\alpha S}{2\sqrt{3}(\alpha\tau+\beta)}\left(\sqrt{3}
\mp \sqrt{1+\gamma} \right), \quad \gamma(\tau) \equiv
\frac{4CS^{2}}{\alpha^{2}}(\alpha\tau+\beta) \label{A0-vel-eq} .
\end{eqnarray}
Introducing another time function $T$ which is also an increasing function of the time $t$, we can rewrite this
equation as follows,
\begin{eqnarray}
\frac{d\gamma}{dT} = \pm \sqrt{1+\gamma}, \quad dT \equiv
\frac{4C}{\sqrt{3}\alpha}S^{2}d\tau=\frac{4C}{\sqrt{3}\alpha S}dt
\label{A0-vel-eq1} .
\end{eqnarray}
Integrating this, we have the solution for the time parameters $T$ as
\begin{eqnarray}
T = T_{0} \pm 2\sqrt{1+\gamma} \label{T-sol} ,
\end{eqnarray}
where $T_{0}$ is an integration constant. Thus, the range for the time $T$ is
determine as
\begin{eqnarray}
T \geq T_{0} +2 ~~{\rm for~the~upper~sign~solution}, \quad  T \leq T_{0} -2
~~{\rm for~the~lower~sign~solution} \label{T-range} .
\end{eqnarray}
From Eq.(\ref{T-sol}), we have a relation between the two time parameters $T$
and $\tau$ as
\begin{eqnarray}
\frac{dT}{d\tau}=\frac{\alpha}{4\sqrt{3}}\frac{(T-T_{0})^{2}-4}{\alpha\tau+\beta}
\label{twotime-rel} .
\end{eqnarray}
Integrating this equation, the solution for the upper signs is obtained as
\begin{eqnarray}
\Phi^{2}=\epsilon\left(\frac{T-T_{0}-2}{T-T_{0}+2}\right)^{\sqrt{3}}, \quad
S(T) =\frac{\alpha}{4\sqrt{\epsilon C
}}\frac{(T-T_{0}+2)^{(\sqrt{3}+1)/2}}{(T-T_{0}-2)^{(\sqrt{3}-1)/2}}
\label{A0-S-sol} ,
\end{eqnarray}
where $\epsilon$ is a positive integration constant, and the integration
constant $T_{0}$ may be chosen as $-2$ such that $T>T_{0}+2=0$. The scale
factor $S(T)$ for the upper signs is a downward concave function in time,
having a minimum as shown in FIG.1.
\begin{figure}[htp]
\includegraphics[width=2.7in]{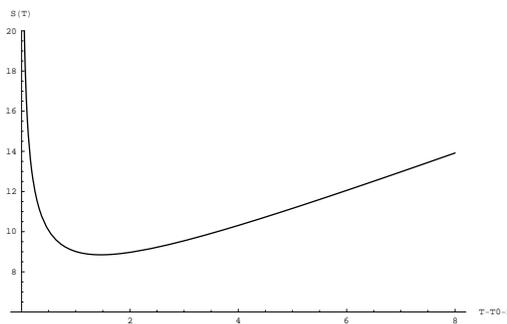}
\caption{$S$ versus $(T-T_{0}-2)$ with $\frac{\alpha}{4\sqrt{C}}=1$ in
Eq.(\ref{A0-S-sol})}
\end{figure}

The cosmological time $t$ is related to the time parameter $T$ as follows
\begin{eqnarray}
t = \int^{T}_{T_{0}+2}\frac{\sqrt{3}\alpha^{2}}{16\sqrt{\epsilon
C^{3}}}\frac{(T-T_{0}+2)^{(\sqrt{3}+1)/2}}{(T-T_{0}-2)^{(\sqrt{3}-1)/2}}dT
\label{time-eq} .
\end{eqnarray}
Integrating this equation, it is found that the cosmological time $t$ is a monotonically increasing function of the
time parameter $T$ with $t=0$ at $T=T_{0}+2$ as shown in FIG.2.
\begin{figure}[htp]
\includegraphics[width=2.7in]{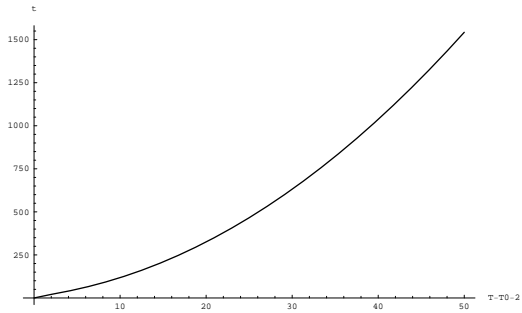}
\caption{$t$ versus $(T-T_{0}-2)$ from Eq.(\ref{time-eq})}
\end{figure}

For the upper sign bounce solution (\ref{A0-S-sol}), as the time $T$ approaches
to $T_{0}+2$, the dilaton field vanishes and the scale factor grows infinitely.
Therefore, the gravitational constant grows infinitely as the cosmological time
approaches to the initial cosmological time $t=0$.

We can examine the cosmological evolution with the help of Eq.(\ref{CosG11-eq}) also. Using the solution
(\ref{phi-vac-sol}) of $\Phi$ and Eq.(\ref{A0-vel-eq}), we find the acceleration of the scale factor from
Eq.(\ref{CosG11-eq}) as
\begin{eqnarray}
\frac{d^{2}S}{dt^{2}} = -\frac{\alpha^{2}}{6S^{5}(\alpha\tau+\beta)^{2}} \left(1 + \frac{\gamma}{2} \mp
\sqrt{3+3\gamma} \right) \label{A0-acc-eq} .
\end{eqnarray}

For the lower signs in Eqs.(\ref{A0-vel-eq},\ref{A0-acc-eq}), the velocity and the acceleration of the scale factor are
both negative always, giving an ever shrinking universe to meet the singularity with the scale factor $S=0$ in the
future.

However taking the upper signs gives the interesting bounce solution as we have obtained in Eq.(\ref{A0-S-sol}). From
Eqs.(\ref{A0-vel-eq},\ref{A0-acc-eq}), the velocity and the acceleration of the scale factor in the time $t$ show the
following behavior,
\begin{eqnarray}
\dot{S} =0 ~~{\rm at} ~~\gamma(\tau)=2, \quad \dot{S} < 0 ~~{\rm for} ~~\gamma(\tau)<2, \quad \dot{S} >0 ~~{\rm for}
~~\gamma(\tau)>2 \nonumber ,~\\ \ddot{S} =0 ~~{\rm at} ~~\gamma(\tau)=4+2\sqrt{6}, \quad \ddot{S} > 0 ~~{\rm for}
~~\gamma(\tau)<4+2\sqrt{6}, \quad \ddot{S} <0 ~~{\rm for} ~~\gamma(\tau)>4+2\sqrt{6} \label{A0-sol} .
\end{eqnarray}
Thus, the scale factor should have a downward concave form in time as seen in FIG.1, having a minimum size at
$\gamma(\tau)=2$. If a universe started its evolution at a finite size with $\gamma(\tau)<2$, it would shrink with a
positive acceleration to the minimum size at $\gamma =2$, which is the turning point. Then, the universe begins an
accelerating expansion again until $\gamma < 4+2\sqrt{6}$, and continues a decelerating expansion afterward. If we
choose $S_{m}$, $\tau_{m}$ as the value satisfying the equation $2CS^{2}_{m}(\alpha\tau_{m}+\beta)=\alpha^{2}$. Then,
the dilaton field $\Phi^{2}$ increases from the value $\Phi_{m}^{2}$ at the minimum size $S_{m}$, and approaches to a
constant value $\sqrt{\epsilon}$ finally in the expanding universe. At the minimum size $S_{m}$, the matter density has
the maximum value $\rho_{max}$ as
\begin{eqnarray}
\rho_{max}=\frac{16\epsilon^{2}C^{3}}{\alpha^{4}}\frac{(\sqrt{3}-1)^{2(\sqrt{3}-1)}}{(\sqrt{3}+1)^{2(\sqrt{3}+1)}},
\quad S_{m}=\frac{\alpha}{2\sqrt{\epsilon C}}\frac{(\sqrt{3}+1)^{(\sqrt{3}+1)/2}}{(\sqrt{3}-1)^{(\sqrt{3}-1)/2}}
\label{minS} .
\end{eqnarray}

Unlike other examples of bouncing scenarios in string cosmology \cite{rev1}, the bouncing solution found with the
radiation-like fluid decoupled from the dilaton field in the string frame does not satisfy the property of duality
symmetry which is typical of string cosmology \cite{rev2}.

\section{The Exact Solutions Both with the 2-form Fields and the Radiation-like Fluid in the String Frame}

Finally, we consider the solutions of Eqs.(\ref{CosG00-eq},\ref{CosG11-eq},\ref{CosD-eq}) with both the radiation-like
fluid and the 2-form fields.

The equation (\ref{CosD-eq}) with the 2-form fields and the radiation-like fluid gives the same solution
(\ref{AB-phi-sol}) of the case with the 2-form fields only. From Eq.(\ref{CosG00-eq}), we have the solution of $H$ as
follows,
\begin{eqnarray}
H=-\frac{\dot{\Phi}}{\Phi} \pm
\frac{1}{\sqrt{3}}\sqrt{\left(\frac{\dot{\Phi}}{\Phi}\right)^{2}+\frac{A^{2}}{4S^{6}}+\frac{C}{S^{4}\Phi^{2}}}
\label{Htau-sol} .
\end{eqnarray}
Using the solution (\ref{AB-phi-sol}) of $\Phi$, from Eq.(\ref{Htau-sol}) we
find the velocity of the cosmic scale factor as follows
\begin{eqnarray}
\frac{dS}{d\tau} = -\frac{AS}{2\sqrt{3}}\csc(A\tau)\left(\sqrt{3}\cos(A\tau)
\mp \sqrt{1+\gamma_{A}} \right), \quad \gamma_{A}(\tau) \equiv
\frac{4CS^{2}}{A^{2}\Phi_{0}^{2}}\sin(A\tau) \label{AB-vel-eq} .
\end{eqnarray}
Similarly as in Eq.(\ref{A0-vel-eq1}), introducing a new time function $T_{A}$,
we can rewrite this equation as
\begin{eqnarray}
\frac{d\gamma_{A}}{dT_{A}} = \pm \sqrt{1+\gamma_{A}}, \quad dT_{A} \equiv
\frac{4C}{\sqrt{3}\Phi_{0}^{2}A}S^{2}d\tau=\frac{4C}{\sqrt{3}\Phi_{0}^{2}A S}dt
\label{Am-vel-eq1} .
\end{eqnarray}
Integrating this, we have the solution for the upper signs as
\begin{eqnarray}
\tan(\frac{A\tau}{2})=\epsilon\left(\frac{T_{A}-T_{0}-2}{T_{A}-T_{0}+2}\right)^{\sqrt{3}},
\quad S(T_{A}) = \frac{\Phi_{0}A}{4\sqrt{2C }} \left(
\epsilon\frac{(T_{A}-T_{0}+2)^{\sqrt{3}+1}}{(T_{A}-T_{0}-2)^{\sqrt{3}-1}}+\frac{1}{\epsilon}\frac{(T_{A}-T_{0}-2)^{\sqrt{3}+1}}{(T_{A}-T_{0}+2)^{\sqrt{3}-1}}
\right)^{1/2} \label{Am-S-sol} ,
\end{eqnarray}
where $\epsilon$ is a positive integration constant, and the integration constant $T_{0}$ may be also chosen as $-2$
such that $T_{A}>T_{0}+2=0$. This scale factor $S(T_{A})$ for the upper signs is a downward concave function in time,
having a minimum size at $\cos(A\tau)=\sqrt{(1+\gamma_{A})/3}$, like the previous upper sign solution for the
relativistic matter only case. Thus, it is found that adding the 2-form fields does not change the qualitative
behaviour of the exact solutions (\ref{A0-S-sol}) for the radiation-like fluid only.

\section{Conclusions and Discussion}

We have discussed a particular class of exact cosmological solutions of the 4-dimensional low energy string gravity
with the 2-form fields and/or the uniform relativistic matter in the string frame (see \cite{intro2} for the general
exact solutions with arbitrary equation of state of the fluid sources). In the vacuum without matter and the 2-form
fields, the exact cosmological solutions always give monotonically shrinking universes if the dilaton field is not a
constant. However, in the presence of a radiation-like fluid and/or the 2-form fields, the exact solutions exhibit a
minimum size of the universe in the string frame. Throughout the evolution, the exact solutions do not show any
singularity, except the initial cosmological curvature singularity in the string frame.

As further studies, in the early stage of cosmological evolution, it may be necessary to consider the effects of the
cosmological constant or a dilaton potential with the radiation-like fluid. Furthermore, in the late stage of the
cosmological evolution, the effects of the dominant non-relativistic fluid should be considered also. Especially, after
the transition from the radiation-dominated to the matter-dominated regime, the qualitative behaviour of the considered
solutions is expected to change in the latest stage of the cosmological evolution, and the dilaton field might be
relevant for phenomenological applications, such as a possible interpretation of the dilaton as the cosmic dark-energy
field \cite{rev3}.

~\\{\it Acknowledgments:} The author would like to thank S.J. Sin for discussions in the early stage of the work and
J.H. Cho for useful comments.




\begin{thebibliography}{99}

\bibitem{IGO1} A. Zee, Phys. Rev. Lett. {\bf 42}, 417 (1979).

\bibitem{IGO2} L. Smolin, Nucl. Phys. {\bf B160}, 253 (1979).

\bibitem{IGO3} A. Adler, Rev. Mod. Phys. {\bf 54}, 729 (1982).

\bibitem{IGO4} A.D. Sakharov. Dokl. Akad. Nauk. SSSR 117, 70 (1967) [Sov. Phys. Dokl. {\bf 12}, 1040
(1967)].

\bibitem{GSW} M.B. Green, J.H. Schwarz, and E. Witten, {\it Superstring Theory}, Vol. I,II (Cambridge University,
1987).

\bibitem{intro} M. Gasperini and G. Veneziano, Astropart. Phys. {\bf 1} 317 (1993).

\bibitem{intro2} M. Gasperini, {\it Elements of String Cosmology}, Cambridege University
Press, Cambridge (2007).

\bibitem{scos1} A.A.Tseytlin, Int. J. Mod. Phys. D {\bf 1}, 223 (1992).

\bibitem{scos2} D.S. Goldwirth and M.J. Perry, Phys. Rev. D {\bf 49}, 5019 (1994).

\bibitem{scos3} E.J. Copeland, A. Lahiri, and D. Wands, Phys. Rev. D {\bf 50}, 4868 (1994).

\bibitem{frame} G. Veneziano, Phys. Lett. B {\bf 406}, 297 (1997).

\bibitem{action} E.S. Fradkin, A.A. Tseytlin, Nucl. Phys. B {\bf 261}, 1 (1985); C.G. Callan, E.J. Martinec, M.J. Perry, and D. Friedan,
Nucl. Phys. B {\bf 262}, 593 (1985); C. Lovelace, Nucl. Phys. B {\bf 273}, 413 (1985).

\bibitem{CY} P. Candelas, G.T. Horowitz, A. Strominger, and E. Witten, Nucl. Phys. {\bf 258}, 46 (1985).

\bibitem{witten} E. Witten, Phys. Lett. B {\bf 155}, 151 (1985); Nucl. Phys. B {\bf 268}, 79 (1986).

\bibitem{string} J. Polchinski, {\it String Theory}, Vols. I and II, Cambridge University Press (1998).

\bibitem{D-brane} C.V. Johnson, {\it D-brane primer}, Boulder 1999, Strings, branes and gravity, 129-350
[arXiv:hep-th/0007170].

\bibitem{gasperini} M. Gasperini and G. Veneziano, Phys. Rep. {\bf 373}, 1-212
(2003).

\bibitem{known1} P.G.O. Freund and M.A. Rubin, Phys. Lett. B {\bf 97}, 233 (1980).

\bibitem{known2} I.C. Rosas-Lopez and Y. Kitazawa, Phys.Rev. D {\bf 82}, 126005 (2010).

\bibitem{rev1} M. Gasperni, M. Giovannini, and G. Veneziano, Nucl. Phys. B {\bf 694}, 206 (2004).

\bibitem{rev2} R. Brustein, M. Gasperini, and G. Veneziano, Phys. Lett. B {\bf 431}, 277 (1988).

\bibitem{rev3} M. Gasperni, Phys. Rev. D {\bf 64}, 043510 (2001).

\end{thebibliography}
\end{document}